\documentclass[epj]{svjour}
\usepackage[dvips]{graphicx}  
\usepackage{amsmath,amsfonts}  
\usepackage{latexsym}
\usepackage{color}
\usepackage{colortbl}
\usepackage{wasysym}
\usepackage{ifsym}
\usepackage{array}
\usepackage{longtable}

\begin{document}

\title{Opinion formation and cyclic dominance in adaptive networks}
\author{G\"{u}ven Demirel\inst{1}  \and  Roshan Prizak\inst{2} \and  P. Nitish Reddy\inst{2} \and Thilo Gross \inst{1} \mail{guven@pks.mpg.de}}
\institute{Max-Planck-Institute for the Physics of Complex Systems -- N\"{o}thnitzer Stra\ss e 38, 01187 Dresden, Germany \and Department of Electrical Engineering, Indian Institute of Technology -- Bombay, Powai, 400076, Mumbai, India  }
\date{Received: \today}

\abstract{
The Rock-Paper-Scissors(RPS) game is a paradigmatic model for cyclic dominance in biological systems. Here we consider this game in the social context of competition between opinions in a networked society. In our model, every agent has an opinion which is drawn from the three choices: rock, paper or scissors. In every timestep a link is selected randomly and the game is played between the nodes connected by the link. The loser either adopts the opinion of the winner or rewires the link. These rules define an adaptive network on which the agent's opinions coevolve with the network topology of social contacts. We show analytically and numerically that nonequilibrium phase transitions occur as a function of the rewiring strength. The transitions separate four distinct phases which differ in the observed dynamics of opinions and topology. In particular, there is one phase where the population settles to an arbitrary consensus opinion. We present a detailed analysis of the corresponding transitions revealing an apparently paradoxial behavior. The system approaches consensus states where they are unstable, whereas other dynamics prevail when the consensus states are stable. 
}

\maketitle

\section{Introduction}
Networks have been used as a metaphor for describing complex systems from a vast range of fields \cite{Albert2002,Newman2003,Newman2006,Boccaletti2006,Dorogovtsev2008,Castellano2009}. 
Many previous studies considered the dynamics \emph{of} networks, a line of research in which the network itself is treated as a dynamical system. 
A prominent example is the preferential attachment mechanism for network growth leading to scale-free topologies \cite{Barabasi1999}. 
Other works focused on the dynamics \emph{on} networks, where each node carries a dynamical state, the time evolution of which is coupled to 
other states according to the network topology. Examples include for instance the synchronization of phase oscillators \cite{Barahona2002,Moreno2004} and epidemics \cite{Pastor-Satorras2001,Brockmann2009} on complex networks.

Until ten years ago the dynamics \emph{of} and \emph{on} networks were studied separately in models, whereas it is clear that both processes typically take place simultaneously in the real world \cite{Gross2008,Gross2009}. 
If one considers both types of dynamics in the same model, an adaptive network is formed \cite{Gross2008,Gross2009}. In the past decade adaptive networks have been studied for instance in the context of epidemiology \cite{Gross2006,Zanette2008,Shaw2008,Funk2010,Marceau2010}, opinion formation \cite{Holme2006,Gil2006,Vazquez2008,Nardini2008,Kimura2008,Benczik2009}, neuroscience \cite{Bornholdt2000,Bornholdt2003,Levina2007,Levina2009,Jost2009,Meisel2009,Ren2010}, and emergence of cooperation \cite{Zimmermann2000,Skyrms2000,Zimmermann2004,Pacheco2006,Poncela2009,Segbroeck2009,Do2010,Zschaler2010}. 
They have been shown to exhibit dynamical phenomena such as robust self-organization to critical behavior \cite{Bornholdt2000}, formation of complex topologies\cite{Holme2006a}, complex system-level dynamics \cite{Ito2001}, and emergence of leadership \cite{Zimmermann2004}. A comprehensive collection of works can be found in \cite{adaptivewiki}. 

An area that has received considerable attention in adaptive networks research is the behavior of simple pa\-ra\-dig\-ma\-tic models of opinion formation. 
These include continuous models such as Deffuant-type compromise models \cite{Kozma2008} and discrete models such as voter-type models \cite{Holme2006,Gil2006,Vazquez2008,Nardini2008,Kimura2008,Benczik2009}. 

The original voter model \cite{Holley1975} considers the spreading of opinions across a static network. The nodes, representing agents in this network, change their opinion dynamically according to a rule capturing social adjustment, the alignment of opinions with neighboring nodes. A second fundamental mechanism not considered in the voter model is social segregation, i.e. preferential linking between agents holding similar opinions and rejection of connections to agents of different opinion. Coupling social segregation with social adjustment leads to an adaptive network. Adaptive voter models \cite{Vazquez2008,Nardini2008,Kimura2008} in which the network topology coevolves with the state of the nodes are especially appealing targets for research, because many results that are found in this simple model appear to hold also in significantly more complex systems in which many different opinions interact by a pairwise-symmetric competition \cite{Holme2006,Kimura2008}. 

For motivating the model considered in the present paper, let us speculate a bit about opinion formation in the real world. In many questions of broad importance such as religious belief or mitigation of climate change, it is evident that there is multitude of different opinions. If pairwise comparisons between opinions are made, there will certainly be some pairings in which the two opinions under consideration are almost equally attractive, 
whereas in other pairings one opinion is clearly much more attractive than the alternative. Previous results \cite{Holme2006,Kimura2008} seem to suggest that even such complex situation can in principle be understood by considering fundamental motifs of relationships between opinions: Those opinions that are clearly inferior to the opinions held by the majority will certainly vanish exponentially. The competition between the remaining opinions, which are almost equally attractive, may then be adequately be captured by the voter-like models. However, one building block is still missing -- the motif of cyclic dominance. This motif depicts a cycle of opinions in which every opinion appears superior in comparison with the next one in the cycle, but inferior when compared to the previous one. 

We believe that cyclic dominance in opinion dynamics can appear when topics of broad interest are discussed in the general public. Although a globally advantageous solution may indeed exist, the agents participating in the opinion formation process are typically non-specialists, which may have only partial knowledge of the situation. For instance in the discussion of public security, it is often quoted that a) additional security measures are necessary to stop an increase in crime, b) closed-circuit cameras are much cheaper than additional police patrols and c) cameras have no quantifyable effect on crime, so one might as well save the expense for the camera. Together, these (admittedly partial) arguments define a cycle of dominance between the three options ``do-nothing'', ``more-police'', and ``more-cameras''.

\begin{figure}[ht!]
  \centering
  \includegraphics[width=1.5in,keepaspectratio]{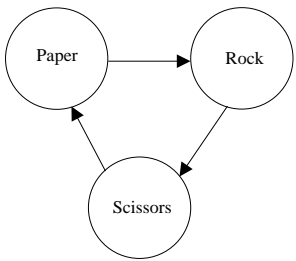}
  \caption{ The cyclic dominance between strategies Rock(R), Paper(P), and Scissors(S). Every option wins against the option its arrow is pointing to, but loses against the third option.}
  \label{fig:game_rule}
\end{figure}

The most simple model of cyclic dominance is known as the rock-paper-scissors (RPS) game. A single RPS game is played by two agents, each choosing between the three pure strategies: rock (R), paper(P), and scissors (S). 
An agent who has chosen R wins against an agent who has chosen S, an agent who has chosen S wins against an agent who has chosen P, and finally an agent who has chosen P wins against an agent who has chosen R. Thus none of the options is globally advantageous (Fig.~\ref{fig:game_rule}). 

The RPS game has been studied in different contexts such as bacterial competition \cite{Iwasa1998,Kerr2002,Kirkup2004}, mating strategies \cite{Sinervo1996,Sinervo2007}, learning in social interaction systems \cite{Sato2002,Sato2003}, and emergence of cooperation \cite{Semmann2003,Hauert2007}. It has been shown that the cyclic dominance functions as the source of biological diversity and species coexistence in many biological systems \cite{Iwasa1998,Kerr2002,Kirkup2004,Sinervo1996,Sinervo2007,Sinervo2006}. For instance, the three-morph mating system in side-blotched lizards displays sustained oscillations, a dynamic form of species coexistence, due to the cyclic dominance relationship between the three morphs \cite{Sinervo1996}. 

The effects of network topology on the RPS game dynamics has been the subject of several previous investigations \cite{Frean2001,Szabo2004,Szolnoki2004,Reichenbach2007,Reichenbach2008,Peltomaki2008}. 
In well-mixed systems, cyclic dominance can be insufficient for ensuring coexistence \cite{Frean2001,Reichenbach2006}, whereas on regular lattices, it is maintained through spiral chaos \cite{Reichenbach2007,Reichenbach2008,Peltomaki2008}. Further, in populations of mobile agents on regular lattices, a non-equilibrium phase transition from coexistence to exclusion is observed as diffusion of agents is increased \cite{Reichenbach2007,Peltomaki2008}. When studied on degree-regular small-world networks with annealed and/or quenched randomness, the RPS game displays non-equilibrium phase transitions between stable coexistence, oscillations, and uniformity \cite{Szabo2004,Szolnoki2004}. 

Here we consider the RPS game on an adaptive social network. Our model captures the cyclic dominance relationship between three opinions and the processes of social adjustment and social segregation.
We start by defining the adaptive RPS game in Section~\ref{sect:model}. Then, we develop a low-di\-men\-si\-o\-nal analytical approximation in Section~\ref{sect:mom-clos-model}. Analytical results from the low-dimensional model and numerical results from agent-based simulations are discussed and compared in Section~\ref{sect:results}. We finish by the summary and discussion of results in Section~\ref{sect:discussion}.


\section{Adaptive RPS game} \label{sect:model}
We consider a network of $N$ nodes and $L$ links, where nodes correspond to agents and links represent social interactions. On the network, each node has an internal state representing its opinion (R, P, or S). The network is initialized as an Erd\H{o}s-Renyi random graph and agents are assigned states randomly with equal probability. The system is then left to evolve according to the following rules: In an update step, a link is chosen at random. If the selected link connects agents in the same state (inert link) then nothing happens. If the link connects agents in different states (active link) then an interaction takes place, from which one agent emerges as the winner and the other as the loser according to the rules of the RPS game. With probability $p$, the loser cuts its connection to the winner and establishes a new connection (rewires) to a randomly chosen agent of its own state. Otherwise (with probability $\bar{p}=1-p$), the loser adopts the opinion of the winner. 

We note that both rewiring and adoption events conserve the total number of nodes and links in the system. Therefore the mean degree $\langle k \rangle = 2L/N$ is time-independent and can be treated as a parameter. 

Throughout this paper we measure time in terms of time steps, corresponding to $N/2$ update steps, such that the expected number of games in which an agent participates per time step is one. We remark that by selecting links, highly connected agents participate in more games. This \emph{link update} rule has been chosen as it presents a compromise between the direct and reverse update rules commonly used in the voter model \cite{Benczik2009}. 


\section{Low-dimensional model} \label{sect:mom-clos-model}
Previous studies showed that the dynamics of discrete adaptive networks can be captured by low-dimensional approximations, describing the global densities of certain subgraphs in the network \cite{Gross2006,Shaw2008,Vazquez2008,Nardini2008,Kimura2008,Zschaler2010,Gross2008a,Risau-Gusman2009}. In the following we consider the densities of nodes of a certain opinion $\rm A\in{\{\rm R, P, S }\} $, denoted by $[A]$, and the densities of links between nodes with given opinions A and B, denoted by $[AB]$. These densities are understood to be normalized to the total number of nodes $N$, such that 
\begin{equation} \label{eq:cons1}
    [R]+[P]+[S]=1 
\end{equation}
and
\begin{equation}\label{eq:cons}
[RR]+[PP]+[SS]+[RS]+[PR]+[SP]=\langle k\rangle / 2.
\end{equation}  

Following the procedure in \cite{Gross2006}, we derive the equations of motion
\begin{equation}\label{eq:ode}\renewcommand{\arraystretch}{2}
\begin{array}{r c l}
\displaystyle \frac{\rm d}{\rm dt} [R]  & = & \displaystyle \bar{p} \left( [RS] - [PR] \right), \\
\displaystyle \frac{\rm d}{\rm dt} [P]  & = & \displaystyle \bar{p} \left( [PR] - [SP] \right), \\
\displaystyle \frac{\rm d}{\rm dt} [S]  & = & \displaystyle \bar{p} \left( [SP] - [RS] \right), \\
\displaystyle \frac{\rm d}{\rm dt} [RR] & = & \displaystyle \bar{p} \left( [RS] + \frac{[RS]^2}{[S]} - \frac{2[PR][RR]}{[R]} \right) + p[PR], \\
\displaystyle \frac{\rm d}{\rm dt} [PP] & = & \displaystyle \bar{p} \left( [PR] + \frac{[PR]^2}{[R]} - \frac{2[SP][PP]}{[P]} \right) + p[SP], \\
\displaystyle \frac{\rm d}{\rm dt} [SS] & = & \displaystyle \bar{p} \left( [SP] + \frac{[SP]^2}{[P]} - \frac{2[RS][SS]}{[S]} \right) + p[RS] \\
\displaystyle \frac{\rm d}{\rm dt} [RS] & = & \displaystyle \bar{p} \left(-[RS] + \frac{2[RS][SS]}{[S]} + \frac{[RP][PS]}{[P]} \right.\\
                           			   &   & \displaystyle         \left.      - \frac{[RS]^2}{[S]} - \frac{[RS][PR]}{[R]} \right)  - p[RS], \\
\displaystyle \frac{\rm d}{\rm dt} [SP] & = & \displaystyle \bar{p} \left(-[SP] + \frac{2[SP][PP]}{[P]} + \frac{[RS][PR]}{[R]} \right.\\
				                   &   & \displaystyle         \left.      - \frac{[SP]^2}{[P]} - \frac{[SP][RS]}{[S]} \right)  - p[SP], \\
\displaystyle \frac{\rm d}{\rm dt} [PR] & = & \displaystyle \bar{p} \left(-[PR] + \frac{2[PR][RR]}{[R]} + \frac{[SP][RS]}{[S]} \right.\\
				                   &   & \displaystyle        \left.      - \frac{[PR]^2}{[R]} - \frac{[PR][SP]}{[P]} \right)  - p[PR] .\\
\end{array} 
\end{equation}

The first three equations describe change in the density of agents holding a given opinion. For instance the first equation captures the change in the density of agents holding opinion R, $[R]$, which depends on the gain from 
agents of opinion S adopting opinion R at the rate $p[RS]$ and the loss from nodes of opinion R adopting opinion P at the rate $p[PR]$. The next three equations in equation~(\ref{eq:ode}) describe the densities of inert links. Every adoption or rewiring event creates at least one inert link. In addition, more inert links can be created in adoption events if the adopting node has multiple neighbors holding the opinion that is adopted. This creation of additional inert links is captured by the quadratic terms in the equations. Finally, inert links can be destroyed if one of connected agents adopts the opinion of a third agent. The corresponding rates then scale both with the density of inert links of a 
given type, say $[RR]$, and the rate at which a given node of the respective type adopts a different opinion, here $p[PR]/[P]$. A factor of two arises because changing the opinion of either of the two nodes on a inert link is sufficient to turn the inert link into an active link. The last three equations in equation~(\ref{eq:ode}) describe the dynamics of active links, which are affected by the same processes as the inert links. 

\begin{figure}[ht!]
  \centering
  \includegraphics[width=3.0in,keepaspectratio]{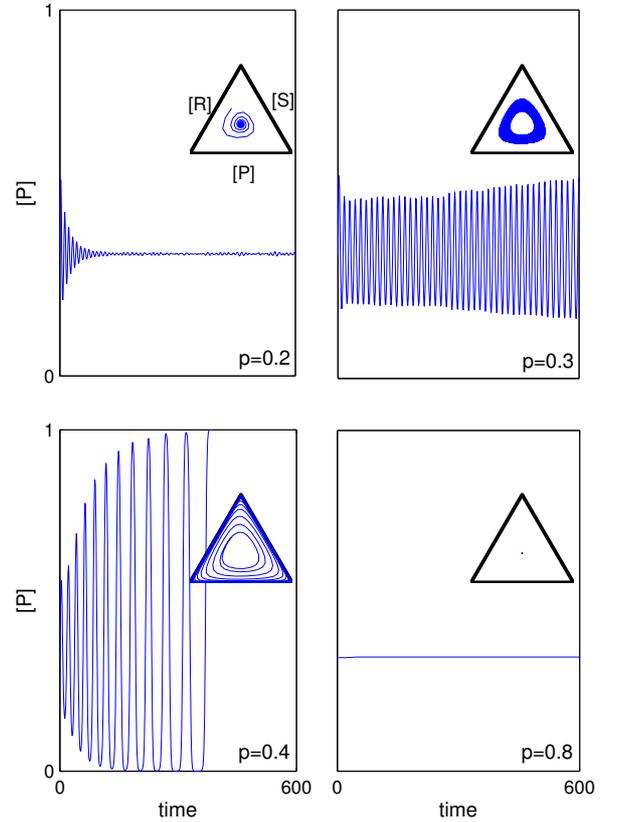}
  \caption{(Color online) Transient and long term dynamics of the RPS system. 
  The timeseries show the density of players holding the opinion P in four different phases: stationary phase (top-left, $p=0.2$), oscillatory phase (top-right, $p=0.3$), consensus phase (bottom-left, $p=0.4$), and fragmented phase (bottom-right, $p=0.8$) The insets show the corresponding long-term behavior in the ternary phase space spanned by the opinion densities R, P, and S. Parameters: $ \langle k \rangle = 4$, $N=10^6$.
  \label{fig:time_evol}}
\end{figure} 

\section{Results} \label{sect:results}
\begin{figure}[ht!]
  \centering
  \includegraphics[width=3.0in,keepaspectratio]{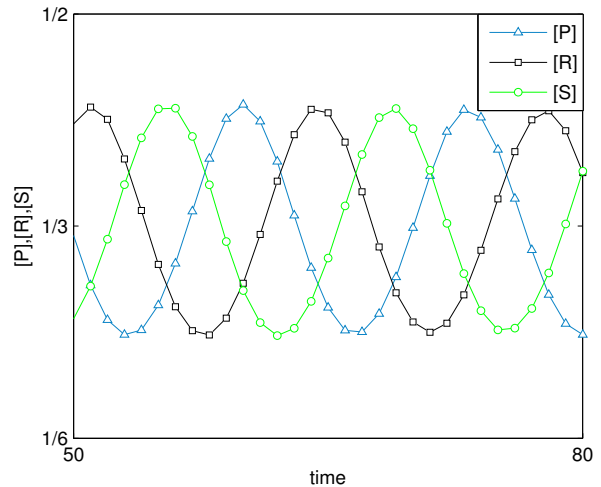}
  \caption{(Color Online) Cyclic succession in the oscillatory phase. Shown are time series of the three variables $[R]$, $[P]$, and $[S]$ in the long-term dynamics. 
  Parameters: $ \langle k \rangle = 4$, $N=10^6$, $p=0.3$.}
  \label{fig:Succession}
\end{figure}

In the following we analyze the dynamics of the RPS game as a function of the parameter $p$, which can be interpreted as the relative frequency of social segregation as response to discomforting interactions.
In representative simulation runs of the agent based model (Figure~\ref{fig:time_evol}), four different types of dynamical behavior are evident: First, at small values of $p$ we observe a \emph{stationary phase}. Although the network remains dynamic on the microscopic level of individual nodes and links, the macroscopic densities of opinions, $[R]$, $[P]$, and $[S]$, approach a state of stationary coexistence in the stationary phase.
Second, at higher values of $p$ there is an \emph{oscillatory phase} where the three opinions show oscillatory behavior as they go through a stable cycle of succession (Figure~\ref{fig:Succession}). Third, if $p$ is increased further then oscillatory behavior is only observed transiently while the system goes through oscillations of increasing amplitude. The system then hits the absorbing boundary where all agents are of the same opinion. In this \emph{consensus phase} all links are inert and the dynamics freezes. Fourth, for high values of $p$ we observe a \emph{fragmented} phase in which rewiring rapidly drives the system to an absorbing state in which the network consists of disconnected communities in which a local consensus is reached. These results are summarized in Table~\ref{tab:phases}

\begin{table}[!ht]
\caption{Dynamical Phases}
\begin{tabular}{|m{1.5cm}|m{1.8cm}|m{1.8cm}|m{1.5cm}|}
\hline \textbf{Phase} &\textbf{Micro  \ \   Dynamics} &\textbf{Macro  \ \ Dynamics} &\textbf{Network}\\ \hline 
Stationary & ongoing & stationary (mixed)& connected \\
Oscillatory & ongoing & oscillatory (mixed)& connected \\
Consensus & frozen & stationary \ \ \ \ \ (polarized) & connected \\
Fragmented & frozen & stationary (mixed) & fragmented \\\hline
\end{tabular}
\label{tab:phases}
\end{table}

\begin{figure}[ht!]
  \centering
  \includegraphics[width=3.0in,keepaspectratio]{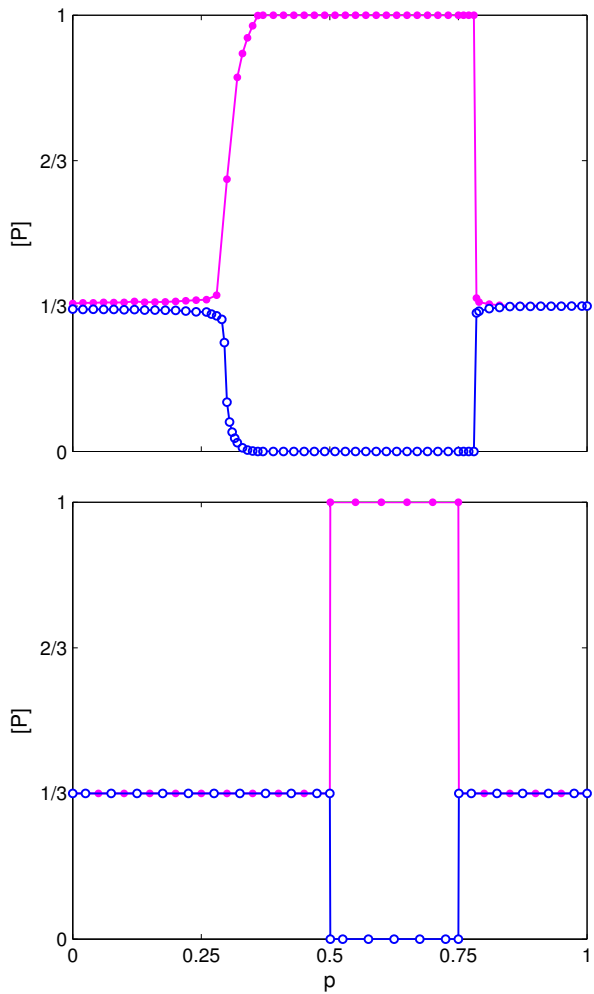}
  \caption{(Color Online) Comparison of numerical results (top) with the analytical approximation (bottom). Shown are the maximum (purple-filled circles) and minimum (blue-empty circles) values of $[P]$ in the long-term dynamics. Both the stationary phase (low $p$) and the fragmented phase (high $p$) are characterized by maxima and minima close to $[P]=1/3$. By contrast in the consensus phase some timeseries approach an all-P state ($[P]=1$) while others approach an all-R or all-S state ($[P]=0$). An oscillatory phase, where $[P]$ oscillates in a finite range, is only observed in the numerical model. Parameters: $ \langle k \rangle = 4$, $N=10^6$. }
  \label{fig:bifurcation}
\end{figure}

\begin{figure}[ht!]
  \centering
  \includegraphics[width=3.0in,keepaspectratio]{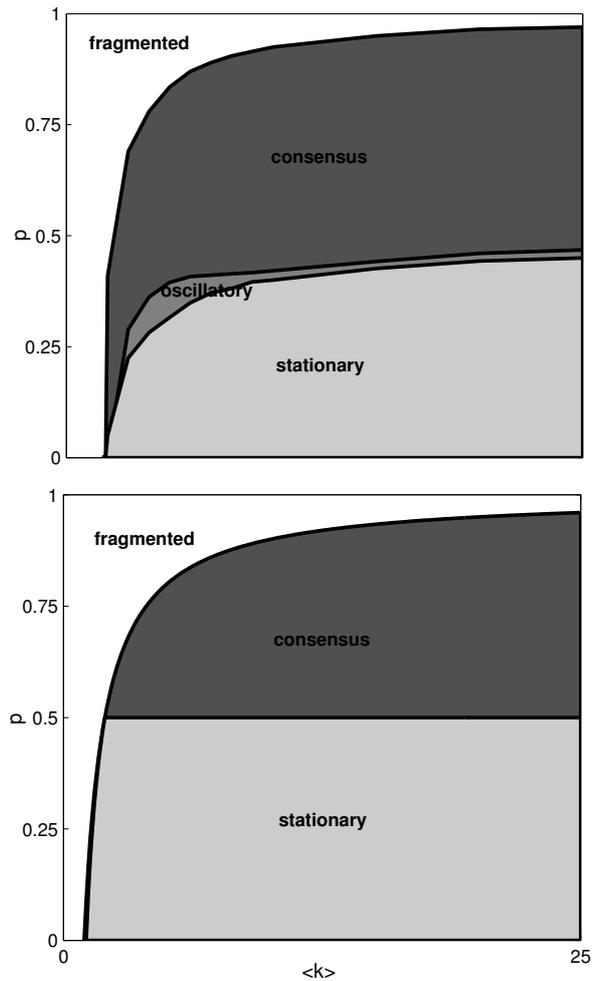}
  \caption{(Color online) Phase diagram of the adaptive RPS game from numerical simulations (top) and the analytical approximation (bottom). 
  Depending on the values of the mean degree $\langle k \rangle$ and the frequency of rewiring $p$ we find four distinct phases (labeled, see also Table~\ref{tab:phases}).
  The oscillatory phase is absent in the analytical approximation. The numerical results are based on networks with $N=10^6$. }
  \label{fig:phase}
\end{figure}

We locate the transitions between the four phases numerically by simulation of the agent-based model and analytically by computation of bifurcations in the low-di\-men\-si\-o\-nal approximation. For illustration, a comparison of the expected maximum and minimum values of $[P]$ in the long-term dynamics is shown in Figure~\ref{fig:bifurcation}. Although the diagrams in Figure~\ref{fig:bifurcation} are similar, there are notable differences: In the analytical model a) 
the oscillatory phase is absent and b) the location of the transition points seems to be relatively poorly approximated. We argue that the analytical approximation is nevertheless a valuable tool. Regarding a) it will become apparent below that the analytical model, despite the absence of the oscillatory phase, provides a conceptual framework for understanding the transitions leading to this pha\-se. Regarding b) we note that the discrepancies arise main\-ly from the relatively low value of $\langle k \rangle$, which we chose specifically to accentuate the differences. The phase diagram in Figure~\ref{fig:phase} shows that for higher $\langle k \rangle$ the numerical results approach the analytical prediction. It is remarkable that for networks with high mean degree $\langle k \rangle$ the oscillatory and fragmented phase occupy only a small portion of the parameter space, such that the system is with high probability in the stationary phase if $p<0.5$ and in the consensus phase otherwise. 
 
For considering the transitions between phases more closely, we start by computing the location of the steady state in which the system resides in the stationary phase 

\begin{equation}\label{eq:activestate}
\begin{array}{r c c c c c l} 
[P]  & = & [R]  & = & [S]  & = & \frac{1}{3}, \\
{[RS]} & = & [SP] & = & [PR] & = & \frac{1}{9}\left(\langle k \rangle-\frac{1}{1-p}\right),\\
{[PP]} & = & [RR] & = & [SS] & = & \frac{1}{9}\left(\frac{\langle k \rangle}{2}+\frac{1}{1-p}\right) .\\
\end{array}
\end{equation}

The corresponding state is stationary irrespective of $p$, but is dynamically unstable for $p>0.5$. Computation of the eigenvalues of the system's Jacobian matrix reveals that the destabilization at $p=0.5$ occurs in a Hopf bifurcation. 
Generically, this type of bifurcation can appear in two different forms \cite{Kuznetsov1998}. In a \emph{supercritical} Hopf bifurcation, a stable limit cycle emerges as the steady state loses stability, whereas in a \emph{subcritical} Hopf bifurcation, an unstable limit cycle contracts around the stable steady state and vanishes as the state is destabilized. For distinguishing between the two forms of Hopf bifurcation one computes the first Lyapunov coefficient in the bifurcation point. This coefficient is negative in the supercritical case and positive in the subcritical case. 

In the present system the corresponding Lyapunov coefficient is exactly zero, which corresponds to a degenerate situation in which the steady forms a center of a dense set of cycles with neutral stability.  

The degeneracy of the Hopf bifurcation in the analytical model is certainly an artefact of the approximation. This problem could be fixed by considering higher order moments in the expansion, which should turn the degenerate bifurcation either into a supercritical or a subcritical Hopf bifurcation. However, already considering network moments of third order increases the number of dynamical equations in the model to 30 and introduces some additional complications \cite{Demirel:InPrep}. In the present paper we therefore do not present the third order expansion, but note that the numerical results strongly suggest a supercritical Hopf bifurcation, which explains the onset of oscillations. 

We note that the location of the Hopf bifurcation chan\-ges at low mean degree. It occurs at $p=(15\langle k \rangle-14-\sqrt{6})/(15\langle k \rangle+2-2\sqrt{6})$ if $(14+\sqrt{6})/15 \leq \langle k \rangle \leq 2$. In this range the Hopf bifurcation is supercritical. However, the band of oscillations is very narrow; they appear only for $(15\langle k \rangle-14-\sqrt{6})/(15\langle k \rangle+2-2\sqrt{6}) \leq p \leq 1-1/ \langle k \rangle$ (Figure \ref{fig:phase}).

Let us now move on to the transition to the consensus phase. In this phase the system approaches one of three consensus states in which all agents on the network hold the same opinion. Consequently, the density of active links is zero ($[RS]=[SP]=[PR]=0$) and the dynamics freezes. Any finite system encountering such an \emph{absorbing state} must therefore remain there for all time. In small systems there is a significant probability that a consensus state is encountered ``by accident'' regardless of the parameter values. However, our numerical results show that in sufficiently large networks the consensus states are only observed in a certain parameter range. For understanding the transitions delimiting this range we therefore consider the thermodynamic limit, $N\to\infty$, captured by the analytical model. 

In the limit of infinite network size, the consensus states are stationary but not absorbing. Because of the normalization, there can be a finite number of active links, even if $[RS]=[SP]=[PR]=0$. Consider a system resting in one of the consensus states, say $[P]=1$, in which a finite number of nodes of state S and a finite number of SP-links remain. If a game is played on one of the SP-links, the agent holding opinion P will lose and react either by rewiring the SP-link or by adopting S. Thus two scenarios are possible, if $p$ is sufficiently large then the removal of SP-links by rewiring events dominates over creation of SP-links by adoption events and the number of SP-links declines exponentially. In this case $[P]=1$ is stable. However, if creation of SP-links by adoption is faster than removal by rewiring then $[SP]$ and $[S]$ grow exponentially. In this case the $[P]=1$ state is a saddle point which has an unstable manifold on which the system can depart toward $[S]=1$.

Intuition suggests that consensus dynamics should be observed only if the consensus states are stable, however, almost the opposite is the case. By stability analysis of the consensus states in the analytical model we find that the consensus state is unstable for $p<p^*=1-1/\langle k \rangle$, showing that consensus states are unstable in the consensus phase. For understanding why these states are observed despite their instability, we return to the example from the previous paragraph. In the parameter range under consideration the $[P]=1$ state is susceptible to the invasion of players holding opinion S. The system will therefore depart exponentially from the $[P]=1$ state and eventually approach $[S]=1$. However, for reasons of symmetry $[S]=1$ must be susceptible to the invasion of R, and $[R]=1$ must be susceptible to the invasion of P. Thus a limit cycle is closed, which connects all three consensus states. Such a cycle connecting multiple saddle-points is called a \emph{heteroclinic loop} \cite{Kuznetsov1998}. Because the approach and departure from a consensus state are exponential, a round-trip on the heteroclinic loop takes infinite time and an observer studying the system on the loop at an arbitrary time will find it in one of the saddle points (here, the consensus states) with probability one. 

For finding the transition marking the onset of the consensus phase we have to ask for the value of $p$ at which not the consensus states, but the heteroclinic loop connecting them becomes dynamically stable. This analysis is complicated by the general difficulties in the stability analysis of limit cycles and the specific degeneracy of our analytical approximation described above. Instead, we therefore settle for a plausibility argument. Consider that the stability of any limit cycle can only change in bifurcations. One such bifurcation is the transcritical bifurcation of cycles, in which two cycles meet and interchange their stability. In the numerical results we have already observed that a stable limit cycle emerges from what we identified as a Hopf bifurcation. When $p$ is increased this cycle grows until it coincides with the heteroclinic loop. At this point the heteroclinic loop becomes stable while the limit cycle is destabilized and leaves the physical space. In the analytical model we do not observe an oscillatory phase, but the dense set of cycles created in the degenerate Hopf bifurcation extends to the heteroclinic loop where it should likewise induce a change in stability.

Finally, we consider the transition from the consensus phase to the fragmented phase. In the analytical model this transition occurs at $p=p^*$, which we identified at the parameter value at which the consensus states become stable. For understanding why the stabilization of the consensus states marks the \emph{end} of the consensus phase, note that the consensus states are not the only absorbing states in the system. To qualify as absorbing, a given state has to satisfy $[RS]=[SP]=[PR]=0$, whereas the other variables, $[R]$, $[P]$, $[S]$, $[RR]$, $[PP]$, $[SS]$, can assume every set of values that is consistent with the normalization. The absorbing states thus include the consensus states and a much larger mass of \emph{fragmented states} in which different opinions survive in disconnected network components. If $p<p^*$ all of these states are dynamically unstable. Because only the consensus states profit from the heteroclinic mechanism, all other do not appear in the long-term dynamics. By contrast, for $p>p^*$ all absorbing states are dynamically stable. The consensus states are now only three points on a huge manifold of stable states and fragmented behavior is observed with high probability. 

In numerical simulations of networks in the fragmented phase we observed that active links are removed so rapidly that only few adoption events occur before the absorbing state is reached. Therefore, the distribution of opinions in the fragmented absorbing state closely mirrors the initial distribution.


\section{Summary and Discussion} \label{sect:discussion}
In the present paper we have proposed a model for the cyclic dominance of three opinions spreading across an adaptive network. An agents $i$ in this network respond to discomforting interactions with another agent $j$ either by social adjustment, adopting $j$'s opinion, or by social segregation, breaking the connection to $j$ and establishing a new connection to a third agent $k$ who shares the opinion of $i$. We showed that the dynamics of this systems depends on the mean degree of the agents and the relative frequency of social segregation. Depending on the values of these parameters there are four distinct dynamical phases, which we characterized as stationary, oscillatory, consensus and fragmented. 

We note that our model combines features that were previously observed in models of cyclic dominance on non-adaptive networks \cite{Sinervo1996,Frean2001,Szabo2004,Szolnoki2004} and in voter-like models on adaptive networks \cite{Vazquez2008,Nardini2008,Kimura2008}. Specifically, behavior closely reminiscent of the stationary, oscillatory and consensus phasea was observed in models of cyclic dominance with different degrees of disorder \cite{Szabo2004} whereas the fragmentation transition, leading to the fragmented phase, was observed in adaptive voter models \cite{Vazquez2008,Kimura2008}. One can therefore suspect that the transitions connecting the stationary, oscillatory, and consensus phase 
could be understood in terms of the disorder that is generated, in our model, intrinsically by rewiring of network connections. By contrast, the fragmentation transition is a genuine adaptive network effect that cannot be observed in non-adaptive networks. In contrast to previous voter-like models, the present system can exhibit sustained long-term dynamics, whereas the voter-like models always reach an absorbing state. 

Our analysis revealed that global consensus is connected to the presence of a heteroclinic loop. A closely related homoclinic mechanism was recently shown to lead to a state of full cooperation in a snowdrift game on adaptive networks \cite{Zschaler2010}. A similar mechanism was also suspected to be at work in the model proposed in \cite{Szolnoki2004}. This evidence suggests that homoclinic and heteroclinic structures could also play a role in other adaptive networks. A more detailed analysis thus appears a promising target for future mathematical research. 

In the introduction we argued that the motif of cyclic dominance could appear in opinion formation processes on questions of broad importance where a population of non-specialists tries to reach a conclusion based on a set of partial arguments (for instance provided by the media). A sensible behavior in this case were to maintain an ongoing discussion until new evidence becomes available that highlights one of the opinions as globally advantageous. 
In our model we find such an ongoing discussion in the stationary phase. In networks of high connectivity, this phase is observed when social adjustment is more frequent than social segregation. However, if segregation dominates then the system most likely ends up in the consensus phase, where all agents agree on one opinion. Which opinion is thus selected is arbitrary, because no opinion is globally advantageous based on the available information. Since this unreasonale response is caused by a collective mechanism we are tempted to call it ``swarm stupidity''. 

If the consensus process, described above, occurred in a real system, the agents would probably be unaware that the decision was made arbitrarily. Because of the cyclic approach to consensus, the consensus decision, say R, has been proven to be superior to the opinion S, which was previously held by the majority of the population. Some agents may remember that earlier S itself replaced P, which was clearly inferior to R (cf.~Figure~\ref{fig:Succession}). Thus looking back from the arbitrary consensus state it may appear that this state was reached by making steady progress from P to S to R. 

Although we are not aware of sociological studies concerning cyclic dominance of opinions, one may speculate on the impact of the internet on opinion formation processes. Personal acquaintances, friends, and family have a high subjective value so the contact to them is likely to be maintained even if they hold a different opinion in certain matters. By contrast it is very easy to ``rewire'' links to online sources of information. It therefore seems natural to assume that an increasing importance of online communication corresponds to an increased frequency of segregation in our model and should therefore promote the ``swarm stupidity'' behavior. 

We believe that in the future the present model may be a building block for investigations focusing on the more realistic scenarios, where there is a complex networks of pairwise dominance relationships between a larger number of opinions.

\end{document}